\documentclass{article}
\usepackage{amsfonts}
\usepackage{amssymb}
\usepackage{epsfig}
\usepackage{dcolumn}
\begin{document}
\title{\textsc{Transport Networks Revisited: Why Dual Graphs?}}
\author{ D. Volchenkov and Ph. Blanchard
\vspace{0.5cm}\\
{\small \it Bielefeld-Bonn Stochastic Research Center (BiBoS)},\\
{\small\it Bielefeld University, Postfach 100131, 33501, Bielefeld, Germany}\\
{\small\it Email: volchenk@physik.uni-bielefeld.de}}

\date{\today}
\maketitle

\begin{abstract}
Deterministic equilibrium flows in transport  networks can be investigated by means
of Markov's processes defined on the dual graph representations of the network.
Sustained movement patterns are generated by a subset of automorphisms
 of the graph spanning the spatial network of a city  naturally
 interpreted as random walks. Random walks assign absolute scores
 to all nodes of a graph and embed space syntax into Euclidean space.
\end{abstract}

\vspace{0.5cm}

\leftline{PACS: 89.65.Lm, 89.75.Fb, 05.40.Fb, 02.10.Ox }

\vspace{0.5cm}

\vspace{0.5cm}


\section{Motivation}
\label{sec:Motivation}
 \noindent

In most of researches
devoted to the improving  of transport routes, the optimization
of power grids, and the pedestrian movement surveys
 the relationships between certain components of the urban texture
  are often measured along streets and routes considered as edges
  of a planar graph, while the traffic end points and street
  junctions are treated as nodes. Such a  {\it primary} graph
  representation of urban networks is grounded on relations
  between junctions through the segments of streets. A usual
   city map based on Euclidean geometry can be considered as an
   example of primary city graphs.

 On the contrary, in another graph representation of urban networks
 known as {\it dual}, relations between streets are encoded through
 their  junctions. Streets within that are treated as nodes while
 their intersections are considered as edges. Dual city graphs are
 extensively investigated within the concept of
{\it space syntax}, a theory developed in the late 1970s, that seeks
to reveal the mutual effects of complex spatial urban networks on
society and vice versa, \cite{Hillier:1984,Hillier:1999}.

Recently, while acting as a referee in {\it Physica A},
 we have been asked to review a manuscript
devoted to city space syntax
in which the dual graph representation of urban patterns is used.
Among the technical remarks expressed by the
previous referees, we have discovered the following
criticism:

\begin{quotation}
\textsl{"[The] {\it dual} option in representing cities would have deserved a
deeper discussion, as it must be said that no one in the world, space
syntax apart, appears to use it ... space syntax
supporters and scientists should dedicate more efforts to the
legitimation of their theory's
fundamental notions in the universally accepted language of the wider
scientific community."}
\end{quotation}

Being not obliged to reply personally, we nevertheless must answer
such a corporative challenge by explaining the deep arguments in
favor of studying namely the dual graph representations.

Legitimacy of the space syntax approach is based on
the surprisingly high
 degree of correlation
between aggregate human movement rates and spatial configuration
analyzed in space syntax theory by means of dual city graphs. We
recall here that the key result of space syntax research is that
the pattern of spatial integration in the urban grid is a key
 determinant of
 pedestrian movement in cities across
  the world, \cite{Hillier2004}.

We emphasize that the dual graph analysis neither
incorporates
many of the factors considered important in previous models of
human behavior in urban environments (such as the motivations and
the origin-destination information) nor direct account was taken
of the metric properties of space (lengthes of streets), \cite{Penn2001}.
Nevertheless,
the robustness of agreement between space syntax predictions
and rush hour movement rates is now supported by a
number  of similar studies of pedestrian movement in different
parts of the world and in an everyday commercial
work of the {\it Space Syntax Ltd.},
 \cite{Read1997}.
Similar results also exist for vehicular movement
\cite{Penn1998} showing that the spatial
configuration of the
urban grid is by itself a
consistent factor driving
transport flows.

The high quality agreement with the empirical data obtained from
pedestrian movement surveys looks amazing even for those who
invented the space syntax approach. Quoting \cite{Penn2001}, there
is nothing else to evoke but that
\begin{quote}
\textsl{"the way people understand their environment and decide on
movement behaviors is somehow implicitly embedded in space syntax
analysis".}
\end{quote}
In the next section, we address three questions of essential importance
 for the analysis of transport networks:
\begin{enumerate}
\item
Why does the space syntax approach based on the pure configurational analysis
 of dual graphs
appear to be so successful in prediction of pedestrian and vehicular flows,
 essentially in rush hours?
\item
Why does the
measurement of the simultaneous
 importance of spaces in a city network by a quantity which is nothing
else, but an element of a transition probability  matrix
determining a Markov chain fit best the empirical data on crowding patterns?
\item
What is the geometrical structure beyond space syntax?

\end{enumerate}

We show that equilibrium transport flows through edges of a connected
undirected graph $G$ at equilibrium are related to the stationary
distribution of random walks defined on its dual counterpart $G^\star.$
Random walks constitute a fundamental process defined on a graph
generated by the set of its automorphisms preserving the notion of
connectivity which
assign the absolute scores to all nodes of the graphs.
 Random walks embed the space syntax of any transport network into
  the Euclidean space $\mathbb{R}^{N-1}$ in which distances and
   angles get clear statistical interpretations.

\section{Mathematics of transport networks}
\label{Arguments}
\noindent

Any graph representation  naturally arises as the outcome of a categorization,
when we abstract a real world system by eliminating all but one of its features
 and by  grouping together things (or places) sharing a common attribute.
For instance, the common attribute of all open spaces
 in city space syntax is that we can move through them.
All elements called
 nodes that fall into one and the same group $V$ are considered as essentially
 identical; permutations of them within the  group are of no consequence.
The symmetric group $\mathbb{S}_{N}$ consisting of all permutations of $N$
elements
($N$ being the cardinality of the set $V$) constitute therefore the symmetry group of $V$.
If we denote by $E\subseteq V\times V$ the set of ordered pairs of nodes called
edges, then  a graph is a map $G(V,E): E \to K\subseteq\mathbb{R}_{\,+}$
(we suppose that the graph has no multiple edges). If two nodes are adjacent, $(i,j)\in E$ we write $i\sim j$.

\subsection{Why dual graphs?}
\label{subsec:1}
 \noindent

First, we establish a connection between transport flows on the graph $G$
and random walks on its dual counterpart $G^\star.$

Given a connected undirected graph $G(V,E)$, in which $V$ is the
set of nodes and $E$ is the set of edges, we introduce the traffic
function $f: E\to(0,\infty[$ through every edge $e\in E$. It then follows
  from the Perron-Frobenius theorem \cite{PerronFrobenius} that the linear equation
\begin{equation}
\label{Lim_equilibrium}
f(e)\,=\, \sum_{e'\sim\, e}\,f(e')\,\exp\left(\,-h\,\ell\left(e'\right)\,\right),
\end{equation}
where the sum is taken over all edges $e'\in E$ which have a common node with $e$,
has a unique positive solution $f(e)>0$, for every edge $e\in E$, for a
fixed positive constant $h>0$ and a chosen set of positive  {\it metric
 length} distances $\ell(e)>0$. This solution is naturally identified
 with the traffic equilibrium state of the transport network defined on
  $G$, in which the permeability of edges depends upon their lengths.
The parameter $h$  is called the volume entropy of the graph $G$, while
 the volume of $G$ is defined as the sum
\[
\mathrm{Vol}(G)\,=\,\frac 12\,\sum_{e\,\in\, E}\,\ell(e).
\]
The volume entropy $h$ is defined to be the exponential growth of the balls in
a universal covering tree of $G$ with the lifted metric, \cite{Manning}-\cite{Lim:2005}.

The degree of a node $i\in V$ is the number of its neighbors in
$G$, $\deg_G(i)=k_i$. It has been shown in \cite{Lim:2005} that
among all undirected connected graphs of normalized volume,
$\mathrm{Vol}(G)=1$, which are not cycles and for which $k_i\ne 1$ for all
nodes,
 the minimal  value of the volume entropy,
$\min(h)=\frac 12\sum_{i\in V}k_i\,\log\left(k_i-1\right)$  is attained
for the length distances
\begin{equation}
\label{ell_min}
\ell(e)\,=\,\frac {\log\left(\left(k_i-1\right)
\left(k_j-1\right)\right)}{2\,\min(h)},
\end{equation}
where $k_i$ and $k_j$ are the degrees of the nodes linked by $e \in E$.
It is then obvious that substituting (\ref{ell_min})
and $\min(h)$ into (\ref{Lim_equilibrium}) the
operator $\exp\left(-h \ell(e')\right)$ is given by
 a symmetric Markov transition operator,
\begin{equation}
\label{Markov_transition}
f(e)\,=\, \sum_{e'\,\sim\, e}\,\frac{f(e')}{\sqrt{\left(k_{i}-1\right)
\left(k_{j}-1\right)}},
\end{equation}
where $i$ and $j$ are the  nodes linked by $e' \in E$, and the sum in
(\ref{Markov_transition}) is taken over all
edges $e'\in E$ which share a node with $e$.
The symmetric operator (\ref{Markov_transition})
 rather describes time reversible random walks over edges than over
nodes.
In other words, we are invited to consider random walks described
by the symmetric operator defined on the dual graph $G^\star$.

The Markov process (\ref{Markov_transition}) represents the
conservation of the traffic volume through the transport network,
while other solutions of (\ref{Lim_equilibrium}) are related to
the possible termination of travels along edges. If we denote the
number of neighbor edges the edge $e\in E$ has in the dual graph
$G^\star$ as $q_e=\deg_{G^\star}(e)$, then the simple substitution
shows that $w(e)=\sqrt{q_e}$ defines an eigenvector of the
symmetric Markov transition operator defined over the edges $E$
with eigenvalue 1. This eigenvector is positive and being properly
normalized determines the relative traffic volume through $e\in E$
at equilibrium.

Eq.(\ref{Markov_transition}) shows the essential role Markov's
 chains defined on edges play in equilibrium traffic modelling and emphasizes that
 the degrees of nodes are a key determinant of the transport networks properties.

The notion of traffic equilibrium had been introduced by J.G. Wardrop in
 \cite{Wardrop:1952} and then generalized in \cite{Beckmann:1956} to a
 fundamental concept of   network equilibrium. Wardrop's traffic
 equilibrium  is strongly tied to the human apprehension of space since
  it is required that all travellers have enough knowledge of the
  transport network they use. The human perception of places is not
  an entirely Euclidean one, but are rather related to the perceiving
  of the vista spaces (viewable spaces of streets and squares) as single units and to the
   understanding of the topological relationships between these vista spaces,
   \cite{Kuipers}.

The use of Eq.(\ref{Markov_transition}) also helps to clarify the inconsistency
of the  traditional axial technique widely implemented in space syntax theory.
Lines of sight are
 oversensitive to small
deformations of the grid, which leads to noticeably different axial graphs for systems
that should have similar configuration properties.
Long straight paths, represented by single lines, appear to be overvalued
compared to curved or sinuous paths as they are broken into a
number of axial lines that
creates an artificial differentiation between straight and curved or sinuous
 paths that have the same importance in the system \cite{Ratti}.
 Eq.(\ref{Markov_transition}) shows that the nodes of a
dual graph representing the  open spaces in the
spatial network of an urban environment should have an individual
meaning being an entity characterized by the certain traffic volume capacity.

Decomposition of city space into a complete set of
   intersecting open spaces characterized by the
traffic volume capacities produces a spatial network which we call
   the {\it dual} graph representation of a city.

\subsection{Why random walks?}
\label{subsec:2}
 \noindent

While analyzing a graph, whether it is primary or dual, we assign
the absolute scores to all nodes based on their properties with
respect to a transport process defined on that. Indeed, the nodes
of $G(V,E)$ can be weighted with respect to some  measure
 $m=\sum_{i\in V}\, m_i \,\delta_{ij},$ specified by a set of positive numbers $m_i> 0$.
The space $\ell^{\,2}(m)$ of square-summable functions with respect to the
   measure $m$ is a  Hilbert space $\mathcal{H}(V)$.

Among all measures which can be defined on $V$, the set of
normalized measures (or {\it densities}),
\begin{equation}
\label{denisty}
1\,=\,\sum_{i\in V}\, \pi_i\,\delta_{ij},
\end{equation}
 are of essential interest since they express the conservation of a
  quantity, and therefore may be relevant to a physical process.

The fundamental physical process defined on the graph is generated by the subset of its
automorphisms preserving the notion of connectivity of nodes. An automorphism is a mapping of the object to itself  preserving all of its structure. The set of all automorphisms of a graph forms a group, called the automorphism group.
For each graph $G(V,E)$, there  exists a unique,
up to permutations of rows and columns,
adjacency matrix $\mathbf{A}$,
 the $N\times N$ matrix defined by
 $A_{ij}=1$ if
  $i\sim j$,  and $A_{ij}=0$ otherwise.
As usual ${\bf A}$ is identified with a linear endomorphism of $C_0(G)$,
the vector space of all functions from $V$ into $\mathbb{R}$.
The degree of a node $i\in V$ is
therefore equal to
\begin{equation}
\label{condition}
k_i\,=\,\sum_{i\sim j}\,A_{ij}.
\end{equation}
Let us consider the set of all linear transformations
defined on the adjacency matrix,
\begin{equation}
\label{lin_fun}
Z\left({\bf A}\right)_{ij}\,=
\,\sum_{\,s,l=1}^N\, \mathcal{F}_{ijsl}\,A_{sl}, \quad \mathcal{F}_{ijsl}\,\in \,\mathbb{R},
\end{equation}
generated by the subset of automorphisms of the graph $G$.

The graph automorphisms are specified by the symmetric group $\mathbb{S}_N$
including all admissible permutations $p\in \mathbb{S}_N$
taking $i\in V$ to $p(i)\in V$. The representation of $\mathbb{S}_N$ consists of all
  $N\times N$ matrices ${\bf \Pi}_p,$ such that
  $\left({\bf \Pi}_p\right)_{i,\,p(i)}=1$, and $\left({\bf \Pi}_p\right)_{i,j}=0$
if $j\ne p(i).$

The function $Z\left({\bf A}\right)_{ij}$ should satisfy
\begin{equation}
\label{permut_invar}
{\bf \Pi}_p^\top\, Z\left({\bf A}\right)\,{\bf \Pi}_p\,=\,
Z\left({\bf \Pi}_p^\top\,{\bf A}\,{\bf \Pi}_p\right),
\end{equation}
for any $p\in \mathbb{S}_N$, and therefore entries of the tensor $\mathcal{F}$
must have the following symmetry property,
\begin{equation}
\label{symmetry}
\mathcal{F}_{p(i)\,p(j)\,p(s)\,p(l)}\,=\,
\mathcal{F}_{ijsl},
\end{equation}
for any $p\in \mathbb{S}_N$.  Since the action of the symmetric
group $\mathbb{S}_N$ preserves the conjugate classes of index
partition structures, any appropriate tensor $\mathcal{F}$
satisfying (\ref{symmetry}) can be expressed as a linear
combination of the following tensors: $ \left\{1,
\delta_{ij},\delta_{is},\delta_{il},\delta_{js},\delta_{jl},\delta_{sl},
\delta_{ij}\delta_{js},\delta_{js}\delta_{sl},\delta_{sl}\delta_{li},
\delta_{li}\delta_{ij},\right.\\
\left.
\delta_{ij}\delta_{sl},\delta_{is}\delta_{jl},\delta_{il}\delta_{js},
\delta_{ij}\delta_{il}\delta_{is} \right\}. $ Given a simple,
undirected graph $G$ such that $A_{ii}=0$ for any $i\in V$ then by
 substituting the above tensors into (\ref{lin_fun}) and taking
 account on symmetries we conclude that any arbitrary linear permutation invariant
 function must be of  the form
\begin{equation}
\label{lin_fun2}
Z\left({\bf A}\right)_{ij}\,=\,a_1+\delta_{ij}\,\left(a_2+a_3k_j\right)+
a_4\,A{}_{ij},
\end{equation}
with  $k_j=\deg_G(j)$ and $a_{1,2,3,4}$ arbitrary constants.

If we require that the
  linear function $Z$ preserves the notion of connectivity,
\begin{equation}
\label{conn_nodes}
k_i\,=\,\sum_{j\in V}\,Z\left({\bf A}\right)_{ij},
\end{equation}
it is clear that we should take $a_1=a_2=0$ (indeed, the
contributions $a_1N$ and $a_2$ are incompatible with
(\ref{conn_nodes})) and then obtain the relation for the remaining
constants, $1-a_3=a_4$. Introducing the new parameter $\beta\equiv
a_4>0$, we write (\ref{lin_fun2}) as follows,
\begin{equation}
\label{lin_fun3}
Z\left({\bf A}\right)_{ij}\,=\, (1-\beta)\,\delta_{ij} k_j +
\beta\,A_{ij}.
\end{equation}
If we express (\ref{conn_nodes})
in the form of the probability conservation relation, then the function
$Z\left({\bf A}\right)$ acquires  a probabilistic interpretation.
Substituting (\ref{lin_fun3}) back into (\ref{conn_nodes}), we obtain
\begin{equation}
\label{property}
\begin{array}{lcl}
1 & = & \sum_{j\in V}\, \frac{Z\left({\bf A}\right)_{ij}}{k_i} \\
  & = & \sum_{j\in V}\, (1-\beta)\,\delta_{ij} +\beta\,\frac{A_{ij}}{k_i}\\
 & = & \sum_{j\in V}\, T^{(\beta)}_{ij}.
\end{array}
\end{equation}
The operator $T^{(\beta)}_{ij}$ represents the generalized random
walk transition operator if $0<\beta\leq k^{-1}_{\max}$ where
$k_{\max}$ is the maximal node degree in the graph $G$. In the
random walks defined by $T^{(\beta)}_{ij}$, a random walker stays
in the initial vertex with probability $1-\beta$, while it moves
to another node randomly chosen among its nearest neighbors with
probability $\beta/k_i$. If we take $\beta=1$, then the operator
$T^{(\beta)}_{ij}$ describes the usual random walks discussed
extensively in the classical surveys \cite{Lovasz}-\cite{Aldous}.

Being defined on a connected aperiodic graph, the matrix $T^{(\beta)}_{ij}$
is a real positive stochastic matrix, and therefore, in accordance to the
 Perron-Frobenius theorem \cite{PerronFrobenius}, its maximal
 eigenvalue is 1, and it is simple. A left eigenvector
\begin{equation}
\label{pi}
\pi\,T^{(\beta)}\,=\,\pi
\end{equation}
 associated with the eigenvalue 1 has positive entries satisfying
  (\ref{denisty}). It is interpreted as a unique
 equilibrium state $\pi$ (stationary distribution of the random walk).
For
  any density $\sigma\in \mathcal{H}(V)$,
\begin{equation}
\label{limit}
\pi\,=\,\lim_{t\to\infty}\,\sigma\, T^{\,t}.
\end{equation}

\subsection{Space syntax as Euclidean space }
\label{subsec:3}
 \noindent

Markov's operators on Hilbert space appear therefore as the natural
 language of complex network theory and space syntax theory in particular.
  Now we demonstrate that random walks embed connected undirected graphs
  into Euclidean space, in which distances and angles acquire the clear
  statistical interpretations.

The Markov operator $\widehat{T}$
is {\it self-adjoint} with respect to the normalized measure (\ref{denisty}) associated
to the stationary distribution of random walks $\pi$,
\begin{equation}
\label{s_a_analogue}
\widehat{T}=\,\frac 12
\left( \pi^{1/2}\,\, T\,\,
\pi^{-1/2}+\pi^{-1/2}\,\, T^\top\,\,
 \pi^{1/2}\right),
\end{equation}
where $T^\top$ is the transposed operator.

In the theory of
  random walks  defined on graphs \cite{Lovasz,Aldous} and in spectral
  graph theory \cite{Chung:1997}, basic properties of graphs are studied in
   connection with the  eigenvalues and eigenvectors of self-adjoint operators
   defined on them.
The orthonormal ordered set of real
 eigenvectors $\psi_i$, $i=1\ldots N$, of the symmetric operator $\widehat{T}$
 defines a basis in  $\mathcal{H}(V)$.

 In particular, the symmetric transition operator  $\widehat{T}$ of the
random walk
 defined on connected
   undirected graphs is
\begin{equation}
\label{transition_symm}
\widehat{T_{ij}}\,=\,
\left\{
\begin{array}{ll}
\frac 1{\sqrt{k_ik_j}},&  i\sim j, \\
0, & i\nsim j.
\end{array}
\right.
\end{equation}
Its first eigenvector
    $\psi_1$ belonging to the largest eigenvalue $\mu_1=1$,
\begin{equation}
\label{psi_1}
\psi_1
\,\widehat{ T}\, =\,
\psi_1,
\quad \psi_{1,i}^2\,=\,\pi_i,
\end{equation}
describes the {\it local} property of nodes (connectivity),
since the stationary distribution of random walks  is
\begin{equation}
\label{stationary}
\pi_i\,=\,\frac{k_i}{2M}
\end{equation}
where $2M=\sum_{i\in V} k_i$.
The remaining eigenvectors,
 $\left\{\,\psi_s\,\right\}_{s=2}^N$, belonging to the eigenvalues
  $1>\mu_2\geq\ldots\mu_N\geq -1$ describe the {\it global} connectedness of the graph.
For example, the eigenvector corresponding to the second eigenvalue $\mu_2$
is used in spectral bisection of graphs. It is called the
Fiedler vector if related to the Laplacian matrix of a graph \cite{Chung:1997}.

Markov's symmetric transition operator $\widehat{T}$  defines a projection
 of any density $\sigma\in \mathcal{H}(V)$ on the eigenvector $\psi_1$ of the
  stationary distribution $\pi$,
\begin{equation}
\label{project}
\sigma\,\widehat{T}\,
=\,\psi_1 + \sigma_\bot\,\widehat{T},\quad \sigma_\bot\,=\,\sigma-\psi_1,
\end{equation}
in which $\sigma_{\bot}$ is the vector belonging to the orthogonal complement of
$\psi_1$.
In space syntax, we are interested in a comparison between the densities  with respect
 to random walks defined on the graph $G$.  Since all components $\psi_{1,i}>0$,
it is convenient to rescale the density $\sigma$ by dividing its
components by the components of $\psi_1$,
\begin{equation}
\label{rescaling}
\widetilde{\sigma_i}, =\,\frac{\sigma_i}{\psi_{1,i}}.
\end{equation}
Thus, it is clear that any two rescaled densities
$\widetilde{\sigma},\widetilde{\rho}\,\in\,\mathcal{H}$ differ
 with respect to random walks only by their dynamical components,
$$
\left(\widetilde{\sigma}-\widetilde{\rho}\right)\,
 \widehat{T}^t\,=\,\left(\widetilde{\sigma}_\bot -\widetilde{\rho}_\bot\right)\,
\widehat{T}^t,
$$
 for all $t>0$.
Therefore, we can define the
distance  $\|\ldots\|_T$ between any two densities established by random walks by
\begin{equation}
\label{distance}
\left\|\,\sigma-\rho\,\right\|^2_T\, =
\, \sum_{t\,\geq\, 0}\, \left\langle\, \widetilde{\sigma}_\bot -\widetilde{\rho}_\bot\,\left|\,\widehat{T}^t\,
\right|\, \widetilde{\sigma}_\bot -\widetilde{\rho}_\bot\,\right\rangle.
\end{equation}
 or, using the spectral
representation of $\widehat{T}$,
\begin{equation}
\label{spectral_dist}
\begin{array}{ll}
\left\|\sigma-\rho\right\|^2_T
&
=
\sum_{t\,\geq 0} \sum_{s=2}^N\, \mu^t_s \left\langle
\widetilde{\sigma}^\bot -\widetilde{\rho}^\bot|\psi_s\right\rangle\!\left\langle \psi_s
| \widetilde{\sigma}^\bot -\widetilde{\rho}^\bot\right\rangle
\\
 &
=\sum_{s=2}^N\,\frac{\left\langle\, \widetilde{\sigma}_\bot -\widetilde{\rho}_\bot\,|
\, \psi_s\,\right\rangle\!\left\langle\, \psi_s\,
| \,\widetilde{\sigma}_\bot -\widetilde{\rho}_\bot\,\right\rangle}{\,1\,-\,\mu_s\,},
\end{array}
\end{equation}
where we have used  Dirac's bra-ket notations especially
convenient for working with inner products and
rank-one
operators in Hilbert space.

If we introduce a new inner product for
densities $\sigma,\rho \in\mathcal{H}(V)$
by
\begin{equation}
\label{inner-product}
\left(\,\sigma,\rho\,\right)_{T}
\,= \, \sum_{t\,\geq\, 0}\, \sum_{s=2}^N
\,\frac{\,\left\langle\,  \widetilde{\sigma}_\bot\,|\,\psi_s\,\right\rangle\!
\left\langle\,\psi_s\,|\, \widetilde{\rho}_\bot \right\rangle}{\,1\,-\,\mu_s\,},
\end{equation}
then (\ref{spectral_dist}) is nothing else but
\begin{equation}
\label{spectr-dist2}
\left\|\,\sigma-\rho\,\right\|^2_T\, =
\left\|\,\sigma\,\right\|^2_T +
\left\|\,\rho\,\right\|^2_T  -
2 \left(\,\sigma,\rho\,\right)_T,
\end{equation}
 where
\begin{equation}
\label{sqaured_norm}
\left\|\, \sigma\,\right\|^2_T\,=\,
\,\sum_{s=2}^N \,\frac{\left\langle\,  \widetilde{\sigma}_\bot\,|\,\psi_s\,\right\rangle\!
\left\langle\,\psi_s\,|\, \widetilde{\sigma}_\bot\, \right\rangle}{\,1\,-\,\mu_s\,}
\end{equation}
is the square
of the norm of  $\sigma\,\in\, \mathcal{H}(V)$ with respect to
random walks defined on the graph $G$.

We finish the description of the $(N-1)$-dimensional Euclidean
space structure of $G$ induced by
  random walks by mentioning that
given two densities $\sigma,\rho\,\in\, \mathcal{H}(V),$ the
angle between them can be introduced in the standard way,
\begin{equation}
\label{angle}
\cos \,\angle \left(\rho,\sigma\right)=
\frac{\,\left(\,\sigma,\rho\,\right)_T\,}
{\left\|\,\sigma\,\right\|_T\,\left\|\,\rho\,\right\|_T}.
\end{equation}
Random walks embed connected undirected graphs into the Euclidean
space $\mathbb{R}^{N-1}$. This embedding  can be used in order to compare
 nodes
and to construct
 the optimal coarse-graining
representations.

Namely, in accordance to (\ref{sqaured_norm}), the density $\delta_i$, which
equals 1 at $i\in V$
and zero otherwise,
acquires the norm $\left\|\,\delta_i\,\right\|_T$
associated to random walks defined on $G$.
In the theory of random walks \cite{Lovasz},
 its square,
\begin{equation}
\label{norm_node}
\left\|\,\delta_i\,\right\|_T^2\, =\,\frac 1{\pi_i}\,\sum_{s=2}^N\,
\frac{\,\psi^2_{s,i}\,}{\,1-\mu_s\,},
\end{equation}
 gets
a clear probabilistic interpretation
expressing
the {\it access time} to a target node
quantifying the expected number
of  steps
required for a random walker
to reach the node
$i\in V$ starting from an
arbitrary
node  chosen randomly
among all other
nodes  with respect to
the stationary distribution $\pi$.

The Euclidean distance between any two nodes of the graph $G$
calculated as the distance (\ref{spectral_dist}) between the densities $\delta_i$ and
$\delta_j$
induced by random walks,
\begin{equation}
\label{commute}
K_{i,j}\,=\,\left\|\, \delta_i-\delta_j\,\right\|^2_T,
\end{equation}
quantifies the {\it commute time} in theory of random walks being
equal to the expected number of steps required for a random
walker starting at $i\,\in\, V$ to visit $j\,\in\, V$ and then to
return back to $i$,  \cite{Lovasz}.

It is important to mention that
the cosine of an angle calculated in accordance to
 (\ref{angle}) has the structure of
Pearson's coefficient of linear correlations
 that reveals it's natural
statistical interpretation.
Correlation properties of flows
of random walkers
passing by different paths
 have been remained beyond the scope of
previous  studies devoted to complex
networks and random walks on graphs.
The notion of angle between any two nodes of the
graph arises naturally as soon as we
become interested in
the strength and direction of
a linear relationship between
two random variables,
the flows of random walks moving through them.
If the cosine of an angle (\ref{angle}) is 1
(zero angles),
there is an increasing linear relationship
between the flows of random walks through both nodes.
Otherwise, if it is close to -1 ($\pi$ angle),
  there is
a decreasing linear relationship.
The  correlation is 0 ($\pi/2$ angle)
if the variables are linearly independent.
It is important to mention that
 as usual the correlation between nodes
does not necessary imply a direct causal
relationship (an immediate connection)
between them.

\section{Discussion and Conclusion}
\label{sec:Summary}
 \noindent

In his speech to the meeting in the House of Lords, October 28, 1943,
 Sir Winston Churchill
had requested that the House of Commons bombed-out on the night of May
 10, 1941, be rebuilt exactly as before. \textit{"We shape our buildings,
  and afterwards our buildings shape us,"} he said.

A belief in the influence of  the built environment on humans was
common in architectural and urban thinking for centuries.
There is a tied connection between
physical activity of humans, their mobility  and the
layout of buildings, roads, and other structures that
physically define a community \cite{Report2005}.
Spatial organization of a place has an extremely
important effect on the way people move
through spaces and meet other people by chance \cite{Hillerhanson}.
It has been also shown in \cite{Language} that space structure and its
impact on movement are critical to the link between the built
environment and its social functioning. In particular,  the reduce
of movement in a  spatial pattern is crucial for the decline of
new housing areas. It is well known that the urban layout effects on the spatial
distribution of crime
\cite{Crime} and poverty \cite{Vaughan02}.

Equilibrium flows in transport  networks can be investigated by means of Markov's
processes defined on the dual graph representations of the network.
Sustained movement patterns are generated by a subset of automorphisms of the
graph spanning the spatial network of a city. This process is naturally
 interpreted as random walks. Random walks assign absolute scores to all
  open spaces in the city accordingly to the quality of paths they provide
  to random walkers and embed city space syntax into Euclidean space.

\section*{Acknowledgment}
\label{Acknowledgment}
\noindent

The support from the Volkswagen Foundation (Germany) in the
framework of the project "{\it Network formation rules, random set
graphs and generalized epidemic processes}" is gratefully
acknowledged.


\begin{thebibliography}{000}

\bibitem{Hillier:1984}
Hillier, B. and Hanson, J. 1984 {\it The Social Logic of Space}. Cambridge University Press. ISBN 0-521-36784-0.
 \bibitem{Hillier:1999}
Hillier, B. 1999 {\it Space is the Machine: A Configurational Theory of Architecture}. Cambridge University Press. ISBN 0-521-64528-X.
\bibitem{Hillier2004}
B. Hillier, {\it The common language of space: a way of looking at the
social, economic and environmental functioning of cities on a common basis},
 Bartlett School of Graduate Studies, London (2004).
\bibitem{Penn2001}
A. Penn, {\it Space Syntax and Spatial Cognition. Or, why the axial line?} In: Peponis, J. and Wineman, J. and Bafna, S., (eds). Proc. of the {\it Space Syntax $3^{rd}$ International Symposium}, Georgia Institute of Technology, Atlanta, May 7-11 2001. A. Alfred Taubman College of Architecture and Urban Planning, University of Michigan, Michigan, USA, 11.1-11.17 (2001).
\bibitem{Read1997}
S. Read, {\it Space syntax and the Dutch city}, in {\it Proc. of
the $1^{\mathrm st}$ Int. Space Syntax Symposium} {\bf 1} 02.1-13,
 Space Syntax Ltd., University College London (1997).

\bibitem{Penn1998}
A. Penn, B. Hillier, D. Banister, J. Xu, {\it Environment
and Planning B: Planning and Design}  {\bf 25}, 59-84 (1998).
\bibitem{PerronFrobenius}
R.A. Horn and C.R. Johnson, {\it Matrix Analysis}, Cambridge University Press, 1990 (chapter 8).
\bibitem{Manning}
A. Manning, {\it Ann. of Math.} {\bf 110}, 567-573 (1979).
\bibitem{Bourdon}
M. Bourdon, {\it L'Einseign. Math.,} {\bf 41} (2), 63-102 (1995) (in French).
\bibitem{Roblin}
T. Roblin, {\it Ann. Inst. Fourier} (Grenoble) {\bf 52}, 145-151 (2002) (in French).


\bibitem{Lim:2005}
S. Lim,  {\it Minimal Volume Entropy on Graphs}. Preprint arXiv:math.GR/050621,
 (2005).

\bibitem{Wardrop:1952}
 J.G. Wardrop, {\it Proc. of the Institution of Civil
Engineers} {\bf 1} (2), pp. 325-362 (1952).

\bibitem{Beckmann:1956}
M.J. Beckmann, C.B. McGuire, C.B.  Winsten,
{\it Studies in the Economics of Transportation.}
Yale University Press, New Haven, Connecticut (1956).

\bibitem{Kuipers}
 B. Kuipers, {\it Environment and Behavior}   {\bf 14} (2),
  pp. 202-220 (1982).


\bibitem{Ratti}
C. Ratti,  {\it Environment and Planning B: Planning and
Design}, vol. {\bf 31}, pp. 487-499 (2004).

\bibitem{Lovasz}
 L. Lov\'{a}sz, {\it Bolyai Society Mathematical
 Studies} {\bf 2}: {\it Combinatorics,
Paul Erd\"{o}s is Eighty}, Keszthely (Hungary), p. 1-46 (1993).
\bibitem{LovWinkl}
L. Lov\'{a}sz, P. Winkler, {\it Mixing of Random Walks and Other Diffusions on a Graph}.
Surveys in combinatorics, Stirling, pp. 119–154 (1995); London Math. Soc. Lecture
Note Ser., vol. {\bf 218}, Cambridge Univ. Press.
\bibitem{Aldous}
D.J. Aldous, J.A. Fill,  {\it Reversible Markov Chains and Random Walks on Graphs}.
A book in preparation, available at www.stat.berkeley.edu/aldous/book.html.


\bibitem{Chung:1997}
F. Chung, {\it Lecture notes on spectral graph theory}, AMS Publications Providence (1997).
\bibitem{Report2005}
{\it Does the Built Environment Influence Physical Activity?
 Examining the Evidence} -- Special Report from the USNational
  Academies' Transportation Research Board and US Institute of Medicine {\bf 282} (2005).

\bibitem{Hillerhanson}
B. Hillier, J. Hanson, {\it The Social Logic of Space} (1993, reprint, paperback
edition ed.). Cambridge: Cambridge University Press (1984).
\bibitem{Crime}
{\it UK Home Office Crime Prevention College Conference on What Really
Works on Environmental Crime Prevention}, October 1998. London UK.

\bibitem{Vaughan02}
L. Vaughan, D. Chatford \& O. Sahbaz {\it Space and Exclusion: The
Relationship between physical segregation. economic marginalization
 and povetry in the city}, Paper presented to Fifth Intern. Space
 Syntax Symposium, Delft, Holland (2005).

\bibitem{Language}
B. Hillier, {\it The common language of space: a way of looking at the social,
 economic and environmental functioning of cities on a common basis}, Bartlett
  School of Graduate Studies, London (2004).

\end{thebibliography}
\end{document}